\documentclass[11pt,twoside]{stm}

\begin{document}
\markboth{\sc E.~Jur\v{c}i\v{s}inov\'a, M.~Jur\v{c}i\v{s}in, R.
Remecky, M. Scholtz}{\sc Scaling Regimes of a Passive Vector Field}

\STM

\title{Influence of weak anisotropy on scaling regimes in a model of advected vector field}

\authors{E.~Jur\v{c}i\v{s}inov\'a$^{1,2}$, M.~Jur\v{c}i\v{s}in$^{1,3}$, R.~Remecky$^4$, M.~Scholtz$^4$}

\address{$^{1}$\,Institute of Experimental Physics SAS, Watsonova 47, 04001 Ko\v{s}ice, Slovakia,\\
$^{2}$\,Laboratory of Information Technologies JINR, 141 980 Dubna,
Russia,\\$^{3}$\,Laboratory of Theoretical Physics JINR, 141 980
Dubna, Russia,
\\$^4$\,Department of Physics and Astrophysics, Institute of Physics,
P.J. \v{S}af\'arik University, \\ Park Angelinum 9, 04001
Ko\v{s}ice, Slovakia}
\bigskip

\begin{abstract}
Influence of weak uniaxial small-scale anisotropy on the stability
of inertial-range scaling regimes in a model of a passive transverse
vector field advected by an incompressible turbulent flow is
investigated by means of the field theoretic renormalization group.
Weak anisotropy means that parameters which describe anisotropy are
chosen to be close to zero, therefore in all expressions it is
enough to leave only linear terms in anisotropy parameters.
Turbulent fluctuations of the velocity field are taken to have the
Gaussian statistics with zero mean and defined noise with finite
correlations in time.  It is shown that stability of the
inertial-range scaling regimes in the three-dimensional case is not
destroyed by anisotropy but the corresponding stability of the
two-dimensional system can be destroyed even by the presence of weak
anisotropy. A borderline dimension $d_c$ below which the stability
of the scaling regime is not present is calculated as a function of
anisotropy parameters.
\end{abstract}

\section*{Introduction}

During the last decade models of scalar or vector fields passively
advected by prescribed stochastic environment have played the
central role in the process of understanding of the so-called
anomalous scaling, the term that refers to the possible deviations
from the predictions of the Kolmogorov phenomenological theory
\cite{MonYag75,Frisch95}. Theoretical understanding of the anomalous
scaling in the framework of a microscopic model remains one of the
main unsolved problems in the theory of fully developed turbulence.
On the other hand, it is well known that the breakdown of the
classical Kolmogorov-Obuchov phenomenological theory of fully
developed turbulence \cite{Frisch95} is even more noticeable for
simpler models of passively advected scalar or vector quantity than
for the velocity field itself and, at the same time, the problem of
a passive advection is easier from theoretical point of view (see,
e.g., \cite{FaGaVe01} and references therein).

One of the most suitable approach for studying self-similar scaling
behavior is the method of the field theoretic renormalization group
(RG) \cite{ZinnJustin,Vasiliev} which was widely used in the theory
of critical phenomena. This method can be also used in the theory of
fully developed turbulence and related problems
\cite{Vasiliev,AdAnVa96,AdAnVa99}, e.g., in the problem of a passive
scalar (or vector) field advected by a given stochastic environment.
In \cite{AdAnVa98} the field theoretic RG was first time applied to
the model of a passive scalar advected by a given statistics of
velocity field, namely, to the so-called Kraichnan model
\cite{Kra68}, where a scalar field is advected by a self-similar
white-in-time velocity field. It was shown that within the field
theoretic RG approach the anomalous scaling is related to the
existence of "dangerous" composite operators with negative critical
dimensions in the framework of the operator product expansion (OPE)
\cite{Vasiliev,AdAnVa96,AdAnVa99}. Afterwards, various generalized
descendants of the Kraichnan model, namely, models with inclusion of
large and small scale anisotropy,
compressibility,
and finite correlation time of the velocity field
were studied by the field theoretic approach (see \cite{Antonov06}
and references therein). Moreover, advection of a passive vector
field by the Gaussian self-similar velocity field (with and without
large and small scale anisotropy, pressure, compressibility, and
finite correlation time) has been also investigated and all possible
asymptotic scaling regimes and crossovers among them have been
classified \cite{all1,AdAnRu01,all2,AnHnHoJu03}. General conclusion
is: the anomalous scaling, which is the most important feature of
the Kraichnan rapid-change model, remains valid for all generalized
models.

In \cite{JuJuReSc06} the influence of the small-scale anisotropy on
the infrared (IR) stability of the possible scaling regimes was
investigated in one particular model of a passive vector advected by
a Gaussian velocity field with finite correlation time in the
presence of the small-scale anisotropy, namely the model where the
stretching term is absent (the so-called $A=0$ model, see, e.g,
\cite{AdAnRu01,AnHnHoJu03}). It can be consider as a starting point
for studying of the influence of anisotropy on the anomalous scaling
of the model. But, as for anomalous scaling, this model is a little
bit specific because in contrast to the other models of passive
vector admixture where the anomalous scaling is related to the
composite operators built of the vector field without derivatives
\cite{all2,AnHnHoJu03} in the case under consideration it is related
to the composite operators built solely of the gradients of the
field. This fact radically changes the complexity of the problem
especially in the anisotropic case (see, e.g.,
\cite{AdAnRu01,Novikov06} and references therein). Thus, as it was
stressed in \cite{JuJuReSc06}, it can be consider as a further step
to the nonlinear Navier-Stokes equation problem. Because the problem
is complicated even at the first stage of the RG analysis (see
\cite{JuJuReSc06}), in what follows, we shall return to the problem
of the influence of the small-scale anisotropy on the IR scaling
regimes, namely, we shall try to understand it when the anisotropy
is considered to be weak. It means that all nonlinear terms in
respect to anisotropy parameters in all expressions can be
neglected. In this case, one has explicit expressions for all
quantities (coordinates of fixed points, eigenvalues of a matrix of
the first derivatives, etc.) and the analysis of the possible
scaling regimes can be done analytically. The results of the present
paper will be used in the subsequent investigations of the anomalous
scaling in the model.

\section*{Formulation of the model}

The model of the advection of transverse (solenoidal) passive vector
field ${\bf b} \equiv {\bf b}({\bf x},t)$ is described by the
following stochastic equation
\begin{equation}
\partial_t {\bf b}  =  \nu_0 \Delta {\bf b} - ({\bf v \cdot \nabla}) {\bf b}   + {\bf f}, \label{K-K}
\end{equation}
where $\partial_t\equiv \partial/\partial t$, $\Delta \equiv{\bf
\nabla}^2$ is the Laplace operator, $\nu_0$ is the diffusivity (a
subscript $0$ denotes bare parameters of unrenormalized theory), and
${\bf v} \equiv {\bf v} ({\bf x} ,t)$ is incompressible advecting
velocity field. The vector field  ${\bf f} \equiv {\bf f} ({\bf x}
,t)$ is a transverse Gaussian random (stirring) force with zero mean
and covariance
\begin{equation}
D_{ij}^f \equiv \langle f_i({\bf x},t) f_j({\bf
x^{\prime}},t^{\prime}) \rangle= \delta(t-t^{\prime})C_{ij}({\bf
r}/L), \,\,\,\,\ {\bf r}={\bf x}-{\bf x^{\prime}} \label{cor-b}
\end{equation}
where parentheses $\langle...\rangle$ hereafter denote average over
corresponding statistical ensemble. The noise defined in
(\ref{cor-b}) maintains the steady-state of the system but the
concrete form of the correlator will not be essential in what
follows. The only condition which must be satisfied by the function
$C_{ij}({\bf r}/L)$ is that it must decrease rapidly for $r\equiv
|{\bf r}| \gg L$, where $L$ denotes an integral scale related to the
stirring.

In real problems the velocity field ${\bf v}(x)$ satisfies
Navier-Stokes equation but, in what follows, we shall work with a
simplified model where we suppose that the velocity field obeys a
Gaussian statistics with zero mean and pair correlation function
\begin{eqnarray}
&&\hspace{-1cm} \langle v_i(x) v_j(x^{\prime}) \rangle \equiv
D^v_{ij}(x; x^{\prime})= \int \frac{d^d {\bf k} d
\omega}{(2\pi)^{d+1}} R_{ij}({\bf k}) D^v(\omega,{\bf k})
e^{-i\omega(t-t^{\prime})+ i{\bf k}({\bf x}-{\bf x^{\prime}})},
\label{corv}
\end{eqnarray}
where $d$ is the dimension of the space, ${\bf k}$ is the wave
vector, and $R_{ij}({\bf k})$ is a transverse projector. In our
uniaxial anisotropic case it is taken as (see, e.g., \cite{all2} and
references therein)
\begin{equation}
R_{ij} ({\bf k})  =
\left(1 + \alpha_{1} ({\bf n \cdot k})^2/k^2\right) P_{ij} ({\bf k})
+ \alpha_{2} n_s n_l P_{is} ({\bf k}) P_{jl} ({\bf k})\,,
\label{T-ij}
\end{equation}
where $P_{ij} ({\bf k})\equiv \delta_{ij}-k_i k_j/k^2$ is common
isotropic transverse projector, the unit vector ${\bf n}$ determines
the distinguished direction, and $\alpha_{1}$, $\alpha_{2}$ are
parameters characterizing the anisotropy. From the positiveness of
the correlation tensor $D^v_{ij}$ one finds restrictions on the
values of the above parameters, namely, $\alpha_{1,2}>-1$. The
function $D^v(\omega, {\bf k})$ in (\ref{corv}) is taken in the
following form \cite{AnHnHoJu03}
\begin{equation}
D^v(\omega, k) = \frac{g_0 u_0 \nu_0^3
k^{4-d-2\varepsilon-\eta}}{(i\omega+u_0 \nu_0
k^{2-\eta})(-i\omega+u_0 \nu_0 k^{2-\eta})}, \label{corrvelo}
\end{equation}
where $g_{0}$  plays the role of the coupling constant of the model
(a formal small parameter of the ordinary perturbation theory), the
parameter $u_{0}$ gives the ratio of turnover time of scalar field
and velocity correlation time, and the positive exponents
$\varepsilon$ and $\eta$ are small RG expansion parameters. The
coupling constant $g_{0}$ and the exponent $\varepsilon$ control the
behavior of the equal-time pair correlation function of velocity
field and the parameter $u_{0}$ together with the second exponent
$\eta$ are related to the frequency $\omega\simeq
u_{0}\nu_{0}k^{2-\eta}$ which characterizes the mode $k$. The value
$\varepsilon=4/3$ corresponds to the celebrated Kolmogorov
\char`\"{}two-thirds law\char`\"{} for the spatial statistics of
velocity field, and $\eta=4/3$ corresponds to the Kolmogorov
frequency. Dimensional analysis shows that $g_{0}$ and $u_{0}$,
which we commonly term as charges, are related to the characteristic
ultraviolet (UV) momentum scale $\Lambda$ (or inner length
$l\sim\Lambda^{-1}$) by \begin{equation}
g_{0}\simeq\Lambda^{2\varepsilon},\qquad
u_{0}\simeq\Lambda^{\eta}.\end{equation}

The stochastic problem (\ref{K-K})-(\ref{corv}) can be rewritten in
a field theoretic form with action functional
\cite{ZinnJustin,Vasiliev}
\begin{eqnarray}
S(\Phi)&=&
b_j^{\prime} \left[ \left(-\partial_t - v_i\partial_i+\nu_0\Delta +
\nu_0 \chi_{10} ({\bf n}\cdot{\bf
\partial})^2 \right)\delta_{jk} + n_j \, \nu_0 \left(\chi_{20} \Delta +
\chi_{30} ({\bf n}\cdot{\bf
\partial})^2 \right) n_k\right] b_k
\nonumber
\\ && - \frac{1}{2}
\left( v_i [D_{ij}^v]^{-1} v_j - b_i^{\prime} D^f_{ij} b_j^{\prime}
\right), \label{action2}
\end{eqnarray}
where $D_{ij}^v$ and $D^f_{ij}$ are given in (\ref{corv}) and
(\ref{cor-b}) respectively, ${\bf b}^{\prime}$ is an auxiliary
vector field (see, e.g., \cite{Vasiliev}), and the required
integrations over $x=({\bf x}, t)$ and summations over the vector
indices are implied. In action (\ref{action2}) the terms with new
parameters $\chi_{10},\chi_{20}$, and $\chi_{30}$ are related to the
presence of small-scale anisotropy and they are necessary to make
the model multiplicatively renormalizable \cite{Vasiliev}. Model
(\ref{action2}) corresponds to a standard Feynman diagrammatic
technique (see, e.g., \cite{all2} for details) and the standard
analysis of canonical dimensions then shows which one-irreducible
Green functions can possess UV superficial divergences. Detail
analysis of the RG technique in the model will be given elsewhere.
We stress only that the functional formulation (\ref{action2}) gives
possibility to extract large-scale asymptotic behavior of the
correlation functions after an appropriate renormalization procedure
which is needed to remove UV-divergences.

\section*{Influence of anisotropy on scaling regimes of the model}

\input epsf
   \begin{figure}[t]
     \vspace{-1cm}
       \begin{flushleft}
       \leavevmode
       \epsfxsize=7.5cm
       \epsffile{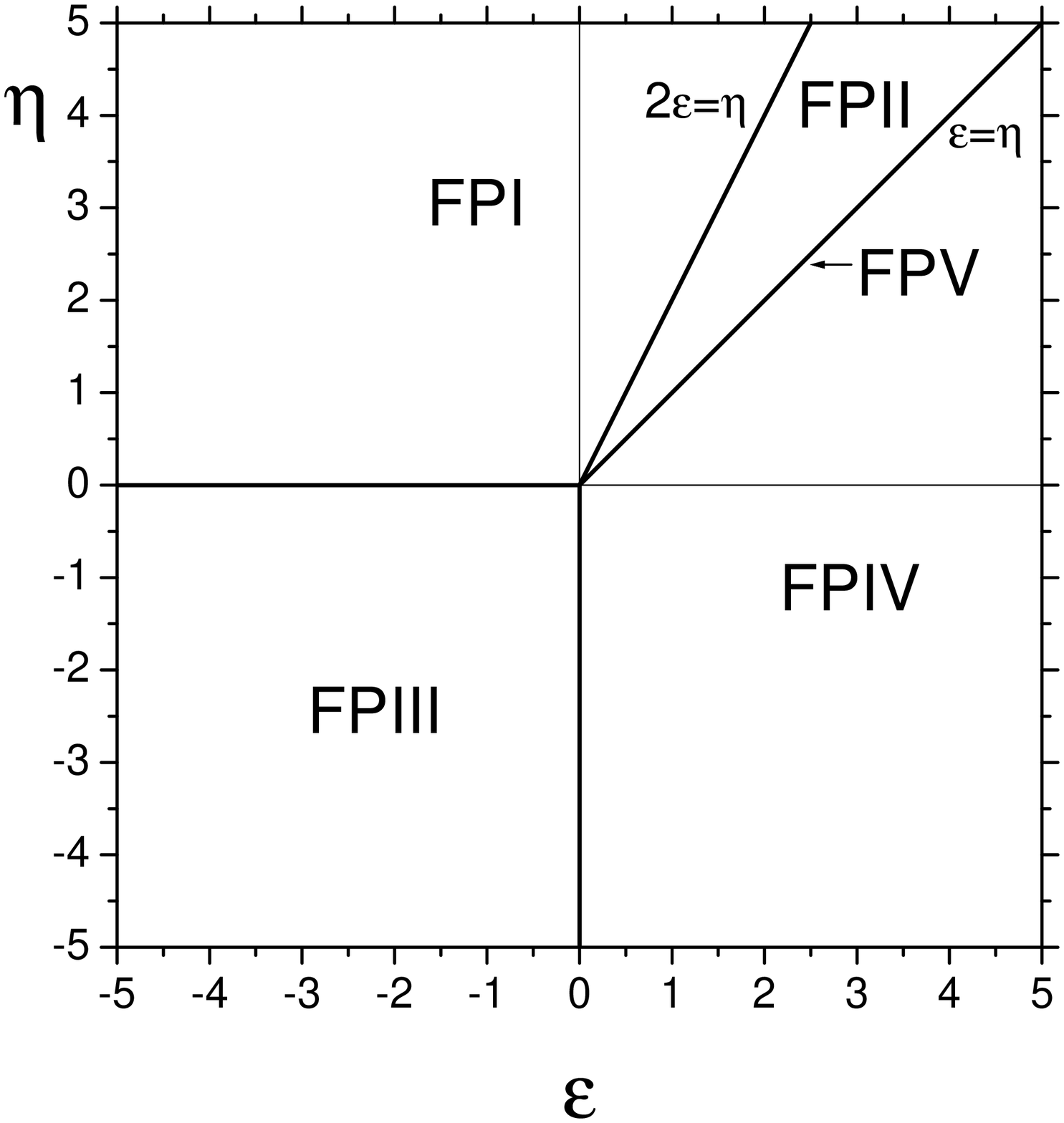}
   \end{flushleft}
     \vspace{-11.7cm}
   \begin{flushright}
       \leavevmode
       \epsfxsize=7.5cm
       \epsffile{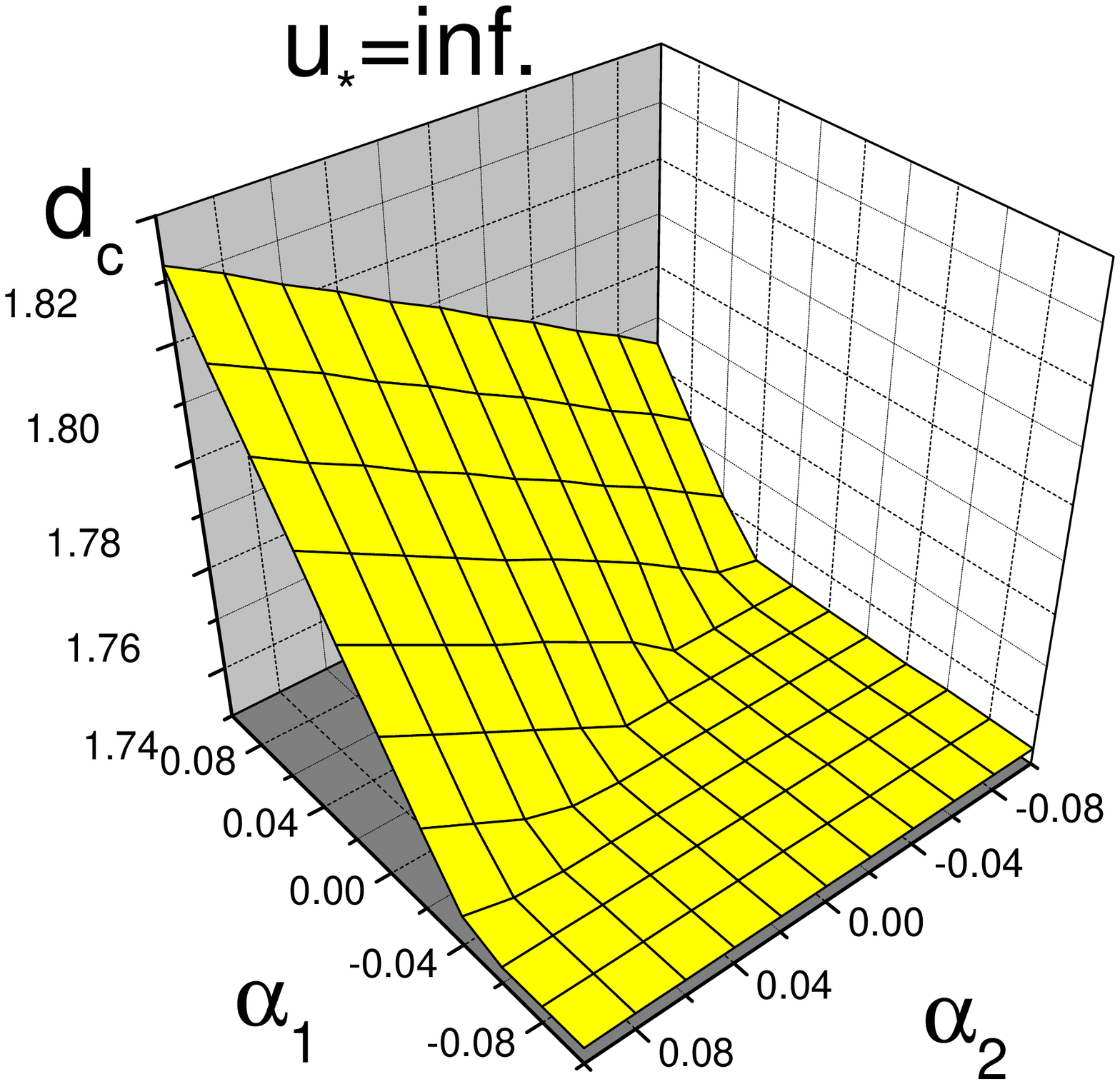}
   \end{flushright}
\vspace{-1.5cm} \caption{(Left) The scaling regimes of the model in
the $\varepsilon-\eta$ plane. The region FPI corresponds to the
trivial rapid-change limit: $g^*/u^*=0, 1/u^*=0$. The region FPII
corresponds to the the non-trivial rapid change limit: $g^*/u^*>0,
1/u^*=0$. The region FPIII corresponds the trivial "frozen" limit:
$g^*=0, u^*=0$. The region FPIV is related to the non-trivial
"frozen" limit: $g^*>0, u^*=0$. In the end, the line
$\varepsilon=\eta$ (FPV) corresponds to the more interesting scaling
regime with $g^*>0, 0<u^*<\infty$. (Right) Dependence of the
borderline dimension $d_c$ on weak anisotropy parameters
$\alpha_{1,2}$ in the Kraichnan limit of the model: $u_*\rightarrow
\infty$. \label{fig1}}
\end{figure}

The  RG analysis leads to the conclusion that possible scaling
regimes are given by the IR stable fixed points of the corresponding
RG equations \cite{Vasiliev,AdAnVa96,AdAnVa99}. The fixed points of
the RG equations can be determined in two ways: First, they are
determined by the corresponding system of RG differential equations
(they are know as the flow equations or Gell-Mann-Low equations),
or, second, they can be found from the requirement  that all the
so-called beta functions of the model vanish and the IR stability of
the fixed point is given by the requirement that all the eigenvalues
of the matrix of the first derivatives $\Omega_{ij}=\partial
\beta_i/\partial C_k$  must have positive real parts, where
$\beta_i$ denotes the full set of beta functions and $C_k$ is the
full set of charges of the model. In the case when uniaxial
anisotropy is strong (nonrestricted), only the first possibility is
suitable. It was briefly discussed in \cite{JuJuReSc06} where the
corresponding analysis of the scaling regimes was done. In what
follows, we shall concentrate on the weak anisotropy limit to better
understand the situation. In this case, the second possibility leads
to the result, i.e., the possible fixed points are given by the
following system of equations
\begin{equation}
\beta_g(g_*,u_*,\chi_{i*},\alpha_j,d,\varepsilon,\eta)=\beta_u(g_*,u_*,\chi_{i*},\alpha_j,d,\varepsilon,\eta)=
\beta_{\chi_i}(g_*,u_*,\chi_{i*},\alpha_j,d,\varepsilon,\eta)=0,\label{betaaaaa}
\end{equation}
for $i=1,2,3; j=1,2$, where each variable with the star denotes the
corresponding fixed point value of the variable and all beta
functions are linear functions of all quantities related to
anisotropy, namely, $\alpha_1, \alpha_2,$ and $\chi_i$, $i=1,2,3$.
The explicit form of the beta functions is as follows
\begin{equation}
\beta_g = g (-2\varepsilon+2\gamma_{1}), \quad \beta_u =
u(-\eta+\gamma_1), \quad \beta_{\chi_i} = \chi_i
(\gamma_1-\gamma_{i+1}),\quad i=1,2,3,\label{betas}
\end{equation}
where, in the limit of weak anisotropy, the so-called gamma
functions $\gamma_i,i=1,2,3,4$ have the following explicit form
\begin{equation}
\gamma_{i+1} = -\frac{g}{2\chi_i} \frac{S_{d}}{(2\pi)^d} \frac{{\cal
K}_{i+1}}{d (d+2)(d+4)(d+6)(1 + u)^2}, \quad i=0,1,2,3\,
\label{gammas2}
\end{equation}
where we define $\chi_0=1$, $S_d=2\pi^{d/2}/\Gamma(d/2)$ is the
$d-$dimensional sphere, and the coefficients ${\cal K}_i, i=1,2,3,4$
are given as follows
\begin{eqnarray}
{\cal K}_1 &=&-(\alpha_2 (3 + d) (6 + d) (1 + u) + \alpha_1 (6 + d)
(1 + d (4 + d)) (1 + u) - 3 (24 + 2 \chi_1 + 6 \chi_2 + 5 \chi_3 +
24 u)  \nonumber \\ && + d (-30 - 25 \chi_1 - 9 \chi_2 - 3 \chi_3 -
30 u + d (21 + 10 d + d^2 - (10+d) \chi_1  - \chi_2 + (3 + d)
(7 + d) u)))\nonumber \\
{\cal K}_2 &=& - 2 (6 + d) ((3 + d) \chi_1 + \chi_2) + 12 \chi_3 - 2
\alpha_1 (3 + d) (6 + d) (1 + u) \nonumber \\&&+ \alpha_2 (6 + d)
(-10 + d (-6 + d (3 + d))) (1 + u)\,,\nonumber \\
{\cal K}_3 &=&  (6 + d) (-2 (3 + d) \chi_1 + (-10 + d (-6 + d (3 +
d))) \chi_2) \nonumber \\ && + (-24 + d (-8 + d (5 + d))) \chi_3 + 2
\alpha_2 (6 + d) (1 + u) + 2 \alpha_1 (3 + d) (6 + d) (1 +
u)\,,\nonumber \\
{\cal K}_4 &=& -2 d (d+4) \chi_3 \,. \nonumber
\end{eqnarray}

\input epsf
   \begin{figure}[t]
     \vspace{-1cm}
       \begin{flushleft}
       \leavevmode
       \epsfxsize=7.5cm
       \epsffile{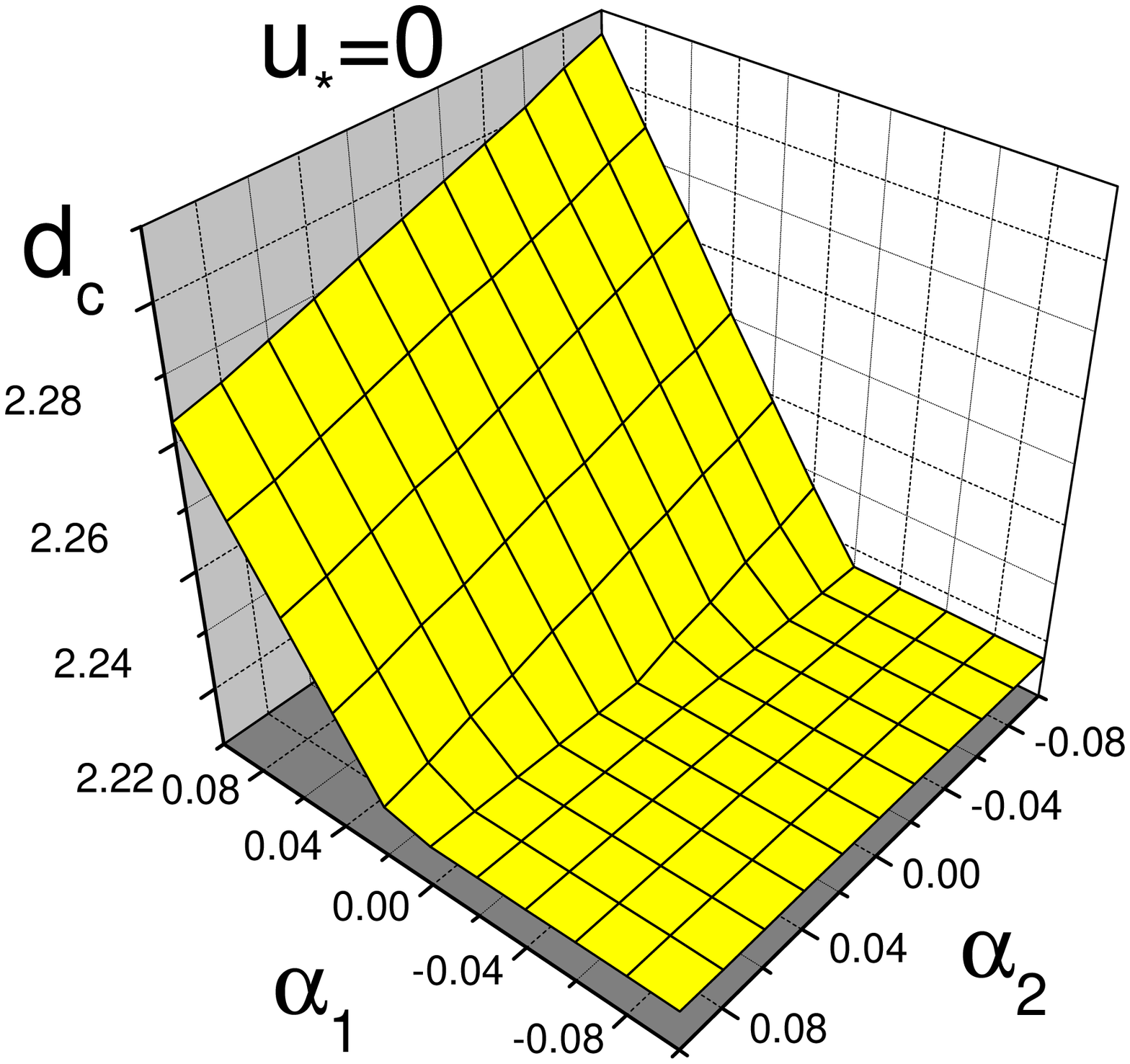}
   \end{flushleft}
     \vspace{-11.7cm}
   \begin{flushright}
       \leavevmode
       \epsfxsize=7.5cm
       \epsffile{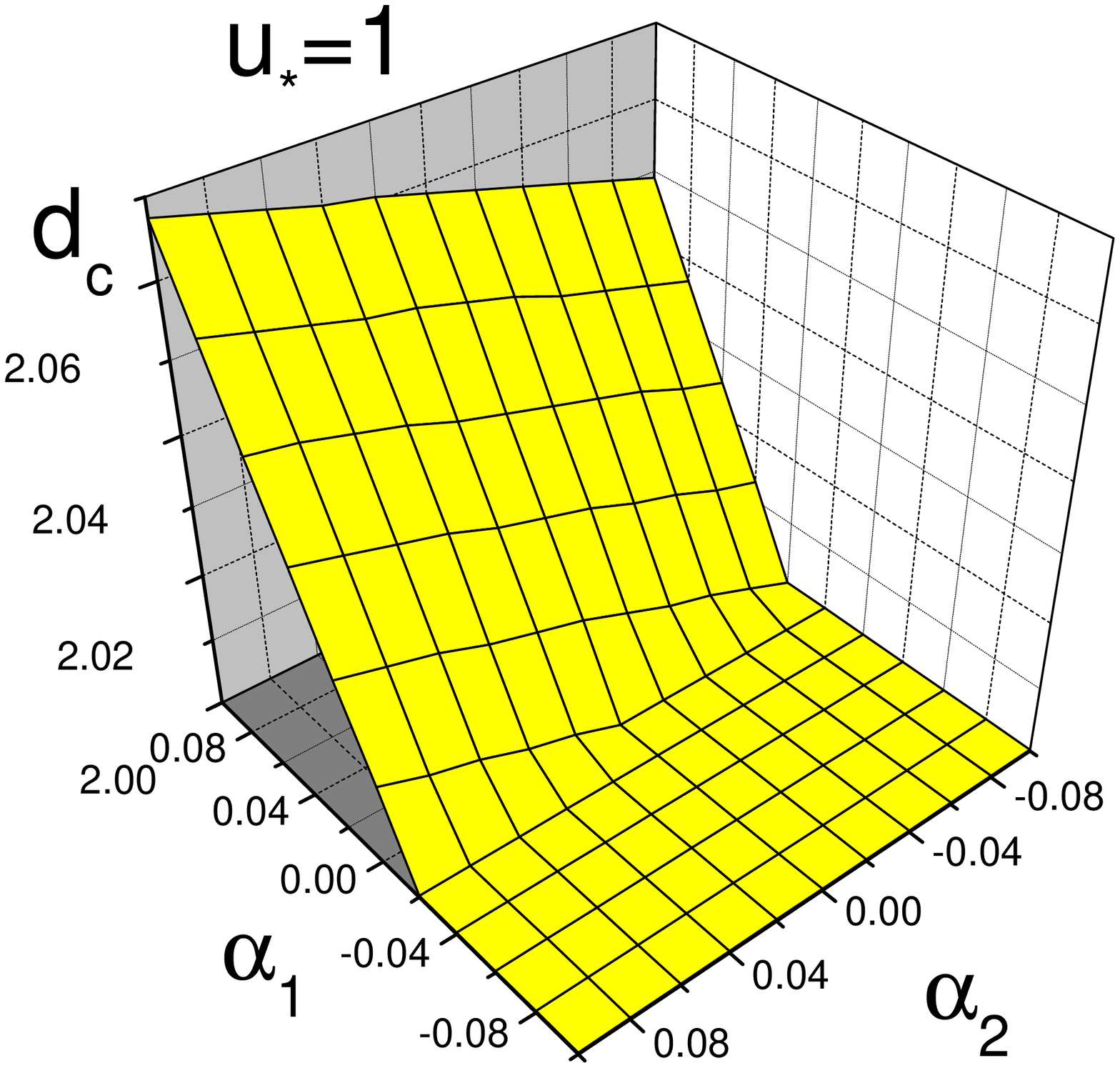}
   \end{flushright}
\vspace{-1.5cm} \caption{Dependence of the borderline dimension
$d_c$ on weak anisotropy parameters $\alpha_{1,2}$ in the case of
the so-called frozen limit of the model: $u_* = 0$, and in the case
with finite value: $u_*=1$. \label{fig2}}
\end{figure}

Thus, possible fixed points are found as solutions of the system of
algebraic equations (\ref{betaaaaa}) and their IR stability is
determined by the positive real parts of the eigenvalues of the
matrix $\Omega=\{\Omega_{ik}\}$. First of all, we have found all
possible IR fixed points which correspond to the possible scaling
regimes and we have analyzed the regions of their IR stability in
the $\varepsilon-\eta$ plane. The results of this analysis are shown
in Fig.\,\ref{fig1}(Left), where it is shown that the model exhibits
five different scaling regimes (two for rapid-change limit, two for
so-called "frozen" limit, and one general with nonzero $u_*$) (see,
e.g., \cite{AnHnHoJu03} and references therein). The same situation
is also held when no restrictions on the uniaxial anisotropy is
supposed \cite{JuJuReSc06}. Further, our interest is concentrated on
the investigation of the dependence of stability of the above
mentioned scaling regimes on anisotropy parameters $\alpha_1,
\alpha_2$ and on the dimension of the space $d$. As in the strong
anisotropy case we have found  the so-called borderline dimension
$d_c$ between stable and unstable regimes as a function of
anisotropy parameters $\alpha_1, \alpha_2$ and parameter $u_*$. The
results are shown in Figs.\,\ref{fig1}(Right) and \ref{fig2} for
different fixed point values of the parameter $u$. The assumption of
linearity of beta functions as a functions of anisotropy parameters
leads to the fact that the borderline surface $d_c=d_c(\alpha_1,
\alpha_2)$ consists of two intersecting planes. One of them is
related to the condition $g_*>0$ and the second surface is related
to the condition to have all real parts of eigenvalues of the matrix
of the first derivatives positive. One can see that the presence of
small-scale anisotropy leads to the violation of the stability of
the corresponding scaling regimes below $d_c\in[2,3]$ for
appropriate values of anisotropy parameters. But from the point of
view of further investigation of anomalous scaling the most
important conclusion is that all the three-dimensional scaling
regimes remain stable under influence of small-scale uniaxial
anisotropy.

\section*{Conclusions}
In present paper we have studied the influence of weak uniaxial
small-scale anisotropy on the stability of the scaling regimes in
the model of a passive vector advected by given stochastic
environment with finite time correlations by means of the field
theoretic RG. The system exhibits five different scaling regimes.
Their IR stability is related to the values of the parameters
$\varepsilon$ and $\eta$.  Besides, the stability of all scaling
regimes is influenced by presence of small-scale anisotropy which is
demonstrated in the existence of the so-called borderline dimension
$d_c$ which is a function of the anisotropy parameters. The $d_c$ is
defined as dimension above which the corresponding scaling regime is
stable and below which the stability of the regime is destroyed. We
have calculated the borderline dimension in the case when anisotropy
parameters are close to zero. The assumption of linearity of beta
functions as a functions of anisotropy parameters leads to the fact
that the borderline surface $d_c=d_c(\alpha_1, \alpha_2)$ consists
of two intersecting planes. All the calculations have been done at
the one-loop level. The results will be used in the further
investigations of the anomalous scaling of the model.

\bigskip

\noindent  ACKNOWLEDGEMENTS --- It is a pleasure to thank the
Organizing Committee of the STM-2006 for kind hospitality. The work
was supported in part by VEGA grant 6193 of Slovak Academy of
Sciences, and by Science and Technology Assistance Agency under
contract No. APVT-51-027904.

\end{document}